\documentclass[preprintnumbers, floatfix, preprintnumbers, letterpaper, twocolumn, superscriptaddress,nofootinbib]{revtex4-2}
\pdfoutput=1 
\usepackage{graphicx}
\usepackage{microtype}
\usepackage{amsmath}
\usepackage{amssymb}
\usepackage{subfigure}
\usepackage{url}
\usepackage{hyperref}
\usepackage{bigints}

\usepackage{xcolor}
\usepackage{color}
\usepackage{mathrsfs}
\usepackage{calrsfs}
\usepackage{amsfonts}
\usepackage{tabularx}
\usepackage{latexsym}
\usepackage{ragged2e}
\usepackage{epsfig}
\usepackage{textcomp}
\usepackage{float}

\usepackage{caption}
\DeclareCaptionJustification{justified}{\leftskip=0pt \rightskip=0pt \parfillskip=0pt plus 1fil}
\definecolor{vividviolet}{rgb}{0.62, 0.0, 1.0}
\definecolor{amaranth}{rgb}{0.9, 0.17, 0.31}
\definecolor{palatinateblue}{rgb}{0.15, 0.23, 0.89}
\definecolor{brightpink}{rgb}{1.0, 0.0, 0.5}
\definecolor{cornflowerblue}{rgb}{0.39, 0.58, 0.93}
\definecolor{deepcarminepink}{rgb}{0.94, 0.19, 0.22}
\definecolor{radicalred}{rgb}{1.0, 0.21, 0.37}

\hypersetup{ linktoc=all,
	colorlinks, linkcolor={palatinateblue},
	citecolor={brightpink}, urlcolor={amaranth}
}

\graphicspath{{Images/}}

\usepackage{tipa}

\def\sideremark#1{\ifvmode\leavevmode\fi\vadjust{\vbox to0pt{\vss% the remark
			\hbox to 0pt{\hskip\hsize\hskip1em%                          will appear only
				\vbox{\hsize1.3cm\tiny\raggedright\pretolerance10000%          on the side
					\noindent #1\hfill}\hss}\vbox to8pt{\vfil}\vss}}}%
\def\beq{\begin{equation}}
\def\eeq{\end{equation}}

\begin{document}
\title{The Case For Black Hole Remnants: A Review}

\author{Yen Chin \surname{Ong}}
\email{ycong@yzu.edu.cn}
\affiliation{Center for Gravitation and Cosmology, College of Physical Science and Technology, Yangzhou University, \\180 Siwangting Road, Yangzhou City, Jiangsu Province  225002, China}
\affiliation{Shanghai Frontier Science Center for Gravitational Wave Detection, School of Aeronautics and Astronautics, Shanghai Jiao Tong University, Shanghai 200240, China}

\begin{abstract}It has been almost 40 years since the proposal of the idea that Hawking radiation of black holes does not lead to a complete evaporation but rather a ``remnant'' state. Though traditionally viewed with great criticisms especially from the high energy physics community, in recent years, various approaches have demonstrated that black hole remnants remain a viable possibility. In this review, which is primarily aimed as an introduction to the subject, we will discuss some possible routes to forming remnants and their respective properties and challenges.
\end{abstract} 

\maketitle
\section{Introduction: The Puzzles of Black Hole Evaporation}

Classical black holes are ``black'' in the sense that nothing, not even light, can escape from inside the horizon. Anything that falls into a black hole is forever lost to the outside world. This naive picture is drastically changed after 
Hawking showed that black holes can radiate quantum mechanically. For an asymptotically flat Schwarzschild black hole, the Hawking temperature\footnote{We will use the units $G=\hbar=c=4\pi\epsilon_0=k_B=1$.} is given by $T=1/8\pi M$. This is inversely proportional to the black hole mass $M$. Therefore, as the black hole loses mass and becomes smaller, it also becomes hotter. The temperature is not bounded above and so diverges in the limit $M \to 0$. The obvious question is: what is the end state of Hawking evaporation? The usual model of Hawking evaporation is governed by the simple ODE of the form
\begin{equation}\label{massloss}
\frac{dM}{dt} = -CAT^4,
\end{equation}
where $C$ is some constant related to the particle species, $A$ the area of the black hole and $T$ the temperature. The right hand side is just the Stefan-Boltzmann law for blackbody emission. The cross sectional area is not simply $A$ as it should be determined by the gray-body factor, and even in the geometric optical limit the length scale  is determined by the impact parameter of the photon sphere instead of just the Schwarzschild radius. Still, modulo these nitty-gritty details, the qualitative feature can be understood by Eq.(\ref{massloss}). Since $T\sim 1/M$ and $A \sim M^2$, we see that $dM/dt \sim -M^{-2}$, which we can solve to find that a black hole whose initial mass is $M$ will evaporate completely in a finite time that scales as $t_\text{life}\sim M^3$. In general relativity, black holes have ``no hair'', which means that there is no other physical parameters that characterize a black hole other than the mass, electric (and hypothetically, magnetic) charge, and angular momentum. Including charge and angular momentum changes the history of the evolution somewhat, but since angular momentum and charge are always radiated away more efficiently than mass loss (in the standard picture of Hawking evaporation), the end state is always a Schwarzschild black hole, so it suffices to consider that as far as remnants are concerned.

A model is of course, only as good as the assumptions we put into it. The above conclusion of a complete evaporation is based on the validity of Eq.(\ref{massloss}), i.e., on the validity of the Stefan-Boltzmann law. Indeed, it is commonly accepted that at sufficiently high energy scale (i.e., high temperature; i.e., small black holes), new physics would likely kick in. This could happen before the Planck scale, and around the Planck scale we would expect quantum gravity effects to become important. Therefore, this already means that the end state of black hole evaporation is not well understood, and on this ground there is no reason to prefer a complete evaporation over the possibility that evaporation somehow effectively stops, becoming a ``remnant''. By remnant, we mean a state of black holes at late time of its evolution, whose lifetime can be infinite, or finite but longer than $M^3$, say $M^4$. 

The aim of this review is to provide an overview of black hole remnants that is accessible to colleagues and students who are familiar with black holes but know little about remnants. There is already a more comprehensive review which I co-authored with Yeom and Chen almost 10 years ago \cite{1412.8366}, and readers are encouraged to refer to that for more details. In this review, I will instead mention some new progress and insights that have been made since then, focusing on (1) how remnants can possibly form and (2) traditional criticisms against remnants and why they do not rule out remnants. Finally we will end with some discussions. 

The possibility of a black hole remnant was proposed already at least as far back as 1987, in the work of Aharonov, Casher, and Nussinov \cite{ACN}, on the ground that Hawking evaporation should satisfy unitarity of quantum evolution\footnote{Essentially, the problem is that if a pure state forms a black hole, while the entirety of quantum state remains pure, the interior and exterior configurations are mixed. Since the interior states presumably cannot escape the black hole, they somehow ``disappear'' if the black hole completely evaporates. This implies that we have a pure-to-mixed evolution, violating unitarity.}. They called remnants ``Planckons''. Indeed, microscopic black holes may look like an ``elementary particle'' at least from the view of an exterior observer (the interior of a small black hole may still be macroscopically large; see below). Some authors have interpreted a black hole to be a collection of many microscopic ``black hole atoms'' \cite{2409.15362}. However, the remnant scenario would be deemed ``impossible'' by later works that instead prefer to maintain unitarity by extracting quantum information from the Hawking radiation itself (see, e.g., the review \cite{1609.04036}). In order to better appreciate the issue, we now give a quick summary of the infamous ``information paradox'' and the ``firewall'' or AMPS paradox \cite{1207.3123,1304.6483} that came after it.

The story started with the realization that black holes have an entropy, characterized by the Bekenstein-Hawking formula $S=A/4$. Black holes need to have entropy otherwise the entropy content of the universe will decrease if we throw things into black holes. The second law of thermodynamics is preserved since the black hole grows afterward. Classically, the area cannot decrease, but when Hawking radiation is included, the \emph{generalized second law} (GSL) still holds: the entropy of the entire system still increases since the emitted Hawking radiation also contains entropy. However, because black holes have no hair, it seems that no information about the stuff that falls into a black hole is retained, or at least not accessible from the observer outside. This is classically acceptable. The problem is when a black hole evaporates -- what happens to the information of everything that has fallen inside? The Hawking radiation is thermal and so does not appear to contain any information. If the black hole eventually completely disappears, then we have lost all the information inside, and this violates unitarity of quantum mechanics (in this context it can be thought of as ``conservation of information''). A possible way out of this conundrum is to allow the information to leak out via Hawking radiation, but in a very subtle manner -- by quantum entanglement. Thus in order to recover the information one has to collect all the Hawking particles and decode the information somehow via a super-quantum computer, impossible in practice but all that is required for unitarity is that this is possible in principle (just like burning a book and recovering the information therein from the ashes is possible in principle, but no one can do it in practice!). 

However, the AMPS paradox pointed out that doing so would lead to another curious issue, namely that at late time (after the black hole entropy has been halved, called the \emph{Page time} \cite{9305007,9306083,1301.4995}), the black hole horizon ceases to be ``uneventful'' and would instead be replaced by a high energy region, dubbed a ``firewall''. Any observer that falls through a black hole in classical general relativity would just pass right through the horizon (a mathematical surface!), but once the firewall sets in, an in-falling observer would now be incinerated at the firewall. The reason for the existence of the firewall is as follows. Hawking particles are formed by pair-production from the vacuum ``near''\footnote{This is not completely correct. For a Schwarzschild black hole the particles can be emitted in a diffused region called ``quantum atmosphere'' that extends some $O(1)$ to $O(10)$ distance away from the horizon \cite{1511.08221}.} the horizon, one of which falls into the black hole while the other is emitted to infinity. The usual argument is that (1) the pair-produced particles are maximally entangled, and that (2) information only starts to be released after Page time. The Hawking particles emitted after Page time needs to purify the earlier Hawking radiation, and therefore they cannot be maximally entangled with their own partner that falls into the black hole. This is due to the monogamy of quantum information: a particle cannot be maximally entangled with two particles at the same time. Without the maximal entanglement between the outgoing and ingoing pairs, the field configuration across the horizon becomes discontinuous. This in turn means that the Hamiltonian (energy) of the field diverges across the horizon (since it contains terms like $\partial^2\phi$, that involves the \emph{derivative} of the field); this is the firewall. An observer that falls into a sufficiently old black hole would encounter the firewall and be destroyed (astrophysical black holes are not yet old enough). Such a drastic departure from the ``no drama'' scenario of general relativity\footnote{The situation is even worse than initially expected. If firewalls can in fact exist, then there is a nonzero probability that a firewall can occur arbitrarily far away from the horizon, a ``naked firewall'' \cite{1511.05695}. Due to the divergence in energy density, a firewall is essentially a singularity; a naked firewall is therefore as bad as a naked singularity. We will have more discussions later on naked  singularity.} led to a fierce debate in the community as to what has possibly gone wrong in the analysis. 

Indeed, there are assumptions that go into the AMPS paradox, and ``resolutions'' of the paradox by challenging these premises can be found throughout the literature. For example, the requirement for maximal entanglement has been challenged \cite{1908.01005,2409.05051}. For our purpose, however, it is the ``central dogma'' \cite{2006.06872} --- the claim that a black hole behaves like a quantum
system whose number of degrees of freedom is proportional to the horizon area --- that requires a closer examination (see, e.g., \cite{2107.05662} for some criticisms of the dogma). The dogma is related to known results that there exists an upper bound for how much information one can squeeze into a finite size region with a given energy, namely the Bekenstein bound \cite{bek} ($S \leqslant 2\pi R E$) and its various more recent generalizations.   
Thus, if the central dogma holds, and if Hawking evaporation does not contain any information, then problems already arise when a black hole becomes sufficiently small (after Page time), regardless of whether there is a complete evaporation at the end. This is because the horizon would eventually be too small to bound the large amount of information inside. In particular this would mean that the remnant picture \emph{for solving the information paradox} is incorrect\footnote{Note that remnants themselves are not incompatible with the central dogma in the sense that if the latter is correct, then a microscopic remnant can only have a tiny amount of information, and therefore cannot solve the information paradox. For the same reason, the firewall would arise as in the standard picture; see \cite{1412.8366}. On the other hand, this also means that if one's goal is to solve the information paradox, it is \emph{not enough} to just come up with a model in which there is a remnant at the end of the evaporation --- one has to assume the failure of the central dogma independently of the existence of the remnant. It is this assumption that is at the heart of the debate.}. However, as we will discuss later on, it could be that the central dogma is not correct, and the remnant picture may still well be viable. But for now, let us begin by explaining how one can obtain remnants in various quantum gravitational models.

\section{How to Obtain Remnants}

A black hole becomes a remnant when it stops evaporating. For example, this can happen if it becomes an extremal black hole. 
In this context, by an extremal black hole we mean the temperature of the black hole is zero (these two notions do not always coincide). 
Note that there is no need for it to become exactly extremal: just by asymptotically approaching an extremal state the black hole evaporation time scale becomes indefinitely long.
Zero temperature extremal black holes are black holes whose inner horizon coincides with the outer horizon. In classical general relativity, one needs to include either a charge or angular momentum in order for a black hole to have two horizons. However, if one envisages removing the black hole singularity, then it is common to end up with ``regular black holes'' with two horizons\footnote{This is not necessarily the case. For example, in \cite{2211.01657}, the authors constructed a regular black hole by considering a massive
source in a quantum superposition of locations. The black hole only has a single horizon, with the zero-temperature extremal case separating the black hole configuration with a (horizonless) wormhole configuration. For another example, see \cite{0902.1746}, in which a single-horizon regular black hole whose singularity is replaced by a bounce was constructed via an effective polymerization scheme in the Loop Quantum Gravity approach.}, even without charge or angular momentum. Such black holes can then radiate indefinitely as they approach the zero temperature limit (whereas for the charged and/or rotating case in general relativity, the charge and angular momentum are typically shed faster than the mass and so the final state is still the Schwarzschild solution). Bardeen black hole \cite{070401}, non-commutative geometry black hole \cite{0510112}, asymptotically safe gravity black hole \cite{1401.4452}, and two-dimensional dilaton black hole \cite{0210016} are just some examples.

Note, however, inner horizons are themselves highly unstable and would likely collapse into another singularity. This seems to defeat the whole point of regularizing the singularity in the first place. Worse, the end state of this instability is unknown, so we cannot say whether there will still be a remnant afterward. One recent progress is to construct black holes with ``inner extremal horizon'', with zero surface gravity, and therefore avoids the classical blueshift instability. Unfortunately this does not seem to avoid the Aretakis instability (classical perturbation does not decay) \cite{1206.6598,1208.1437,1212.2557,2112.09832}. In addition, semi-classically the inner extremal horizon remains unstable \cite{2303.03562}. This way of obtaining a remnant is therefore quite challenging conceptually, but clearly more studies are required, as the properties of inner horizon of a black hole are still unclear in general \cite{2308.09225}, and the possibility that black holes can asymptote towards some extremal configurations and thus become an effective remnant cannot yet be ruled out \cite{2405.08069}. Note that extremal black hole is not the only possible end state, as some regular black holes can evolve into a horizonless object \cite{2211.05817,2302.00028,2410.10582}. 

Another possibility is to consider quantum gravitational effect that kicks in when the black hole is sufficiently small. Note the subtle difference: in the contexts of regular black holes, while the classical singularity could be removed by some kind of quantum gravity effect, remnants are the result of the existence of extremal state. The inner horizon --- which would not be there without the quantum gravitational effect regularizing the singularity --- exists already when the black hole is still astronomically large. Thus, these extremal remnants are not \emph{directly} caused by any quantum gravitational correction, only indirectly so\footnote{Admittedly, what qualifies as ``direct'' is rather subjective.}. A more direct influence from quantum gravity can happen at or near the Planck scale, without necessarily modifying the black hole solution when $M$ is still large, and this can be studied phenomenologically by employing the ``generalized uncertainty principle" (GUP). 

Let us restore $\hbar, c$ and $G$ for the moment for clarity.
The most basic form of GUP is given by\footnote{The notation $\Delta x^n$ is a short hand of $(\Delta x)^n$. Similarly for $\Delta p^n$.}
\begin{equation}\label{GUP}
\Delta x \Delta p \geqslant \frac{1}{2}\left(\hbar + \alpha L_p^2 \frac{\Delta p^2}{\hbar}\right),
\end{equation}
where $L_p$ is the Planck length, and $\alpha$ is a dimensionless parameter, typically taken to be of order unity. We will first discuss the case in which $\alpha > 0$.
From Eq.(\ref{GUP}) we can obtain the following inequality:
\begin{flalign}\label{inequa}
&\frac{\hbar}{\alpha L_p^2}\left[\Delta x \left(1-\sqrt{1-\frac{\alpha L_p^2}{\Delta x^2}}\right)\right] \leqslant \Delta p  \\ \notag
&\leqslant \frac{\hbar}{\alpha L_p^2} \left[\Delta x \left(1+\sqrt{1-\frac{\alpha L_p^2}{ \Delta x^2}}\right)\right].
\end{flalign}

The presence of the square root makes it clear that GUP imposes a minimum uncertainty in the position: 
\begin{equation}\label{xmin}
\Delta x_\text{min} = {\sqrt{\alpha}}{L_p}.
\end{equation} 
Hawking temperature is modified under GUP. For an asymptotically flat Schwarzschild black hole this modification was obtained heuristically in \cite{0106080} by considering the characteristic energy of the Hawking particle as $E = pc$ and by identifying $\Delta x$ with the horizon scale $\Delta x \sim 2GM/c^2$. This gives (upon multiplying with a normalization constant prefactor $1/2\pi$ in order to obtain the correct temperature when $\alpha \to 0$):
\begin{equation}\label{TGUP}
T_\text{GUP} = \frac{Mc^2}{\pi \alpha}\left(1-\sqrt{1-\frac{\alpha \hbar c}{4GM^2}}\right).
\end{equation}
Note that the minus sign before the square root is chosen (instead of the positive one; c.f. Eq.(\ref{inequa})) so that the correct Schwarzschild temperature is obtained in the $\alpha \to 0$ limit.
The problem with such a heuristic treatment is that we cannot be sure of its validity. After all, while the method works for the Schwarzschild case, it does \emph{not} lead to the correct known temperature\footnote{By ``known temperature'' we mean the expression obtained using standard methods like Hawking's original Bogoliubov transformation, the Euclidean Wick rotation method, quantum anomaly calculations, or even tunneling method. They all give the same temperature expression.} for the Kerr or Reissner-Nordstr\"om black hole \cite{2407.21114} (though intriguingly, how it fails may be related to black hole thermodynamics \cite{2407.21114}), so how could we be confident that it works when there is a GUP correction? For other criticisms for heuristic treatments of GUP, see \cite{2005.12075,2303.10719,2305.16193}.

Eq.(\ref{TGUP}), if indeed correct, gives rise to a peculiar behavior that the temperature becomes nonzero but finite at the minimum mass\footnote{This can also be obtained from Eq.(\ref{xmin}) by identifying $\Delta x$ with the horizon scale $2GM/c^2$.} 
\begin{equation}\label{Mmin}
M_\text{min}=\frac{\sqrt{\alpha}M_p}{2}, 
\end{equation}
where $M_p$ denotes the Planck mass. The fact that there is a nonzero temperature yet the black hole no longer evaporates (otherwise the square root term in Eq.(\ref{TGUP}) becomes imaginary)
could be a sign that modifying the Hawking temperature without modifying the Schwarzschild metric may be problematic.
Furthermore, if we look at the temperature of a Reissner-Nordstr\"om black hole in standard general relativity, there is also a square root involved, namely a factor of $\sqrt{M^2-Q^2}$ (in Planck units). In this case the temperature tends to zero as $M \to Q$, so the $M=Q$ state is an extremal black hole. Still, even then, there are considerable interests in the possibility of a naked singularity formation. That is, whether some kind of perturbation can bring $Q$ to exceed $M$. Crucially, note that $\sqrt{M^2-Q^2} \notin \Bbb{R}$ does not imply there is no state such that $Q>M$, only that such a state would no longer be a black hole. Therefore, in the GUP case discussed above, it is actually not necessary that the $T \neq 0$ remnant ``ceases radiating'' because the temperature becomes imaginary. Rather, precisely because $T \neq 0$, there is a risk that it may drive the evolution further beyond a black hole stage into another horizonless configuration, if not a naked singularity. Whether this is the case would require a careful investigation.

Of course, the whole idea of a classical, well-defined, metric tensor might already fail near the Planck scale, so it is not clear if any modified metric is trustworthy (worse still, different metrics can be obtained by relying on different heuristic ``derivations'' \cite{2303.10719}). Nevertheless, we note that the entropy of the black hole, which can be obtained by integrating the first law of black hole thermodynamics, can be shown to satisfy (returning to Planck units hereinafter)
\begin{equation}\label{SGUP}
S_\text{GUP} = 4\pi M^2 - \frac{1}{2}\pi \alpha \ln M + \sum_{n=1}^\infty c_n (4\pi M^2)^{-n} + \text{const.}
\end{equation}
The logarithmic correction term is consistent with many approaches of quantum gravity (though the sign differs from one approach to another), which gives us some confidence that despite the heuristic argument, Eq.(\ref{TGUP}) may be correct.

Surprisingly, the problem of the final non-evaporating but nonzero temperature black hole can be avoided. This is due to another effect of GUP that can prevent black hole evaporation even when the black hole is still astronomically large (and therefore when the temperature remains largely unmodified), just after a scrambling time \cite{2309.01638}. If correct, this renders the aforementioned Planckian size black hole remnant irrelevant except for black holes that were formed to be microscopic in the beginning.
For a Schwarzschild black hole, the scrambling time \cite{0708.4025,0808.2096} is $r_S\log(r_S/L_p)$, where $r_S=2M$ denotes the horizon; it is the time scale for information to be scrambled over the horizon after being dropped into a black hole. Note that this time scale is far shorter than $\sim r_S^3$, the lifetime of a black hole. Thus, if a large black hole stops evaporating after the scrambling time, then it becomes a macroscopic remnant and the peculiarity mentioned above never arises. Intuitively this result can be appreciated as follows. First, we recall the heuristic derivation of Hawking temperature above, in which the Hawking temperature scale is roughly set by $\Delta p$, and by the uncertainty principle $\Delta x \sim 1/\Delta p$, we can trace $\Delta x$ back to the horizon scale, i.e., Hawking radiation comes from the region around the black hole. However, with GUP, it can be shown that beyond the scrambling time, $\Delta p \gg \alpha/L_p^2$, which in turn implies that $\Delta x \gg r_S$, which cannot be interpreted as particle creation by the black hole \cite{2309.01638}. See also \cite{2309.12926}, in which the same conclusion was reached using a string theoretic approach. Note that if correct, this also avoids the firewall problem since the black hole never becomes small enough to get past the Page time scale. The remnant is essentially a \emph{classical} black hole. To contrast with this scrambling time scale remnant,
in another recent approach, it is argued that a black hole stops evaporating after it has emitted about half of its initial mass, due to ``memory burden effect'' \cite{2405.13117}, in which a system with high capacity of information storage (such as a black hole) is stabilized by the load of information it carries. Thus we see that it is not necessarily the case that remnants are microscopic.

It is almost poetic that a quantum gravitational correction like the one in GUP, which we usually think can only be important when the black hole is small, can give rise to a drastic deviation from semi-classical picture even when the black hole is still huge (and even renders the black hole effectively classical with no Hawking radiation). Yet, this is not the first time we have seen such an unexpected behavior. For another example, consider the Chandrasekhar limit of a white dwarf star. White dwarfs are stable because the electron degenerate pressure counteracts gravitational collapse. This pressure is inherently quantum mechanical in nature, it arises from the jittering motion (large $\Delta p$) of the particles inside the star, which increases as the star becomes smaller (small $\Delta x$). Once GUP is taken into account, it seems that the Chandrasekhar limit is destroyed \cite{1512.06356,1804.05176}. Dynamical stability analysis can restore an upper mass limit for white dwarfs for a certain range of the GUP parameter \cite{2002.08360}, which explains why in astrophysical observation we do not observe arbitrarily large white dwarfs. Still, here too we see that a supposedly small quantum gravity effect can lead to big changes in physics well outside the quantum gravity regime.

In \cite{1806.03691} I considered Hawking evaporation under the assumption that the GUP parameter could be negative. In such a scenario there is no longer a minimum mass, and so the black hole can continue to evaporate indefinitely. However, its lifetime turns out to be infinite, so we still have a remnant that asymptotes to zero mass. There is, however, a peculiar property --- the zero mass black hole (or rather, no more black hole!) has a finite nonzero temperature in the negative GUP parameter case. This is as weird, if not weirder than, the positive GUP parameter case in which the non-zero temperature remnant supposedly does not radiate. Fortunately, since this bizarre state cannot be reached in a finite time, it is not physically relevant as the end state of an evaporating black hole\footnote{There is still one subtle issue: a zero mass black hole should be easily produced by quantum fluctuations, so why don't we see them? The reason may have to do with their nonzero temperature, which in a consistent quantum gravity treatment may contribute to a nonzero effective energy-momentum tensor and this then gives it an effective mass after all.}. See also \cite{1812.03136} for more discussions.

So far we have discussed extremal black holes and the GUP approaches that led to remnants. In recent years, another promising possibility has come to light: a \emph{white hole} remnant, which is a scenario supported by loop quantum gravity \cite{1801.03027,1802.04264,1805.03872,1905.07251,2009.05016,2105.06876,2207.06978}, see the recent review \cite{2407.09584} for details (and \cite{2411.03125} for a nice summary). The basic idea is simple: the static Schwarzschild solution does not distinguish the direction of time, and indeed in the maximally extended spacetime we see that there are black hole \emph{and} white hole, described by the same metric tensor. Classically, we do not usually care about the white hole part because it is unphysical (in the sense that being a time-reversed version of black hole, a white hole cannot be produced by gravitational collapse). However, quantum mechanically one should consider a superposition of both black and white holes, and the possible tunneling from the black hole state to the white hole one. This probability becomes significant as the black hole becomes sufficiently small. In this scenario, a black hole with initial mass $M$ evaporates with lifetime that scales like $M^3$ as in the standard Hawking picture. However, before it completely evaporates away, it tunnels into the white hole state, which then becomes a remnant that slowly radiates away. The time scale for this diffuse emission is about $M^4$ \cite{2207.06978}. Such a white hole remnant does not live indefinitely. However, like the more traditional black hole remnant it can conceivably still solve the information paradox, as we will discuss in the next section. 

There are of course other routes that can possibly lead to black hole remnants, but more studies are required to check these claims. For example, many quantum gravity theories predict that at sufficiently high energy the effective dimension of spacetime is 2 instead of 4 (see \cite{1705.05417} for a review; also see \cite{2111.05018} and the references therein). In \cite{1204.3619} it was proposed that if the early 2-dimensional universe is governed by a dilatonic gravity theory, a black hole can exist. However, as the universe expanded and cooled down, the spacetime transited into 3 dimensions (before it becomes 4). In 3-dimensions however, if there is no negative cosmological constant, then there can be no black hole solution \cite{0005129}. Thus, the black holes from the 2-dimensional epoch ceased its evaporation and became ``stuck'' as remnants. This picture is however, somewhat problematic. What is the geometry of a black hole remnant once it crossed the dimensional transitions? Most remnants \emph{are} themselves black holes, so if no black hole solution is allowed, what are those remnants, which in any case, have two dimensions less than the surrounding space? Also, most quantum gravity results indicate that spacetime dimension does not change, but rather the spectral dimension or thermal dimension does (i.e., spacetime is only \emph{effectively} 2-dimensional at high energy scale). In this case there is no problem with black holes with effective dimension being 3. While the scenario above concerns the dimensional transition of the entire universe, we can also consider the dimensional transition of the black hole itself. As the black hole evaporates and gets hotter, eventually its spectral dimension reduces to 2. It is shown in \cite{2111.05018} that provided that the solution still only has 1 horizon, the lifetime is only slightly longer than the standard $M^3$, namely $M^3 + cM$ term, where $c$ is a constant. Therefore it seems that the spectral dimension route does not lead to remnants. However, in the case of \emph{thermal dimension} \cite{1602.08020,2009.08556}, which arises from the modification of the dispersion relation, whether remnants occur or not likely depend on the dispersion relation. Also related to dispersion relation is the idea of ``gravity's rainbow'' \cite{0305055}, which does lead to black hole remnants \cite{1410.5706}. 

\section{Why Traditional Criticisms Against Remnants are Not Convincing}

There are two major issues concerning remnants, which led to their viability being questioned by the mainstream high energy physics community. The two problems are the \emph{species problem}\footnote{Not to be confused with another ``species problem'' \cite{9409015,1206.2365} (though they may be related): if Bekenstein-Hawking entropy is entanglement entropy, it should depend on the species of particles/fields available, so why is it only proportional to the area?} and the \emph{entropy problem}, which are closely related. We begin with the species problem. Consider again the no-hair theorem of black holes. Once a black hole is formed, there is no way to tell from the outside what was the stuff that collapsed to form the black hole in the first place. The possibility seems to be endless, each of these different internal states gives rise to a distinct ``species''. Now, consider the probability of pair-producing remnants from quantum fluctuation of the vacuum or in a high energy experiment such as a collider. For each species, whose mass is about the Planck mass, the pair-production probability is tiny. However, since there are infinitely many species, the total production rate seems to be unbounded. This therefore seems to predict that tiny black holes should be copiously produced from the vacuum all the time, as long as the available energy exceeds twice the remnant mass (given that they are Planckian, this is a relatively low energy requirement). The fact that we do not observe them therefore seemingly rules out black hole remnants. This criticism is, however, over-simplistic, as we do not know if the number of internal states is actually infinite. In fact, Giddings had argued that it is finite \cite{9310101,9412159}. Even if there is indeed an infinitely many possible internal states, the usual physics of \emph{particle} pair-production may be quite misleading when applied to remnants. Even though they may appear particle-like with mass the order of the Planck mass, they still have a vast interior structure. The vast amount of space inside a black hole is a result that follows from standard general relativity, not any exotic physics. 

For any star of radius $r$, given that it is spherical, we can deduce that its volume has to be fixed, given by $\frac{4}{3}\pi r^3$. However, a black hole is a highly curved spacetime manifold; its interior is nontrivial. This should be well-known but somehow the literature is still confusticated by statements about the ``central singularity'' of a Schwarzschild black hole, as if $r=0$ is at the geometric center of a sphere\footnote{The ``$r=\text{const.}$'' surfaces inside the black hole are sometimes referred to as ``T-spheres'' of the ``T-region'', especially in older texts \cite{ruban} to emphasize the distinction.}. What happens in the Schwarzschild case is that the singularity is spacelike, an event in the future of the in-falling observer. The interior spacetime is best viewed as a special case of Kantowski-Sachs cosmological spacetime, which is anisotropic \cite{0609042, brehme}. In particular, inside the horizon, a line of constant $r, \theta, \phi$ has a proper length that goes to infinity as the singularity is approached, hence ``there is no dearth of space inside'' of a black hole, in the words of Ted Jacobson \cite{0308048}.

A notion of interior volume, independent of coordinates, was defined by Christodoulou and Rovelli \cite{1411.2854}, which for a Schwarzschild black hole, is best computed in 
the Eddington-Finkelstein coordinates,
grows linearly in advanced time $v$ as\
\begin{flalign}
\text{Vol.} &\sim \int^v \int_{S^2}  \max\left(r^2 \sqrt{\frac{2M}{r}-1}\right)  \sin\theta ~d\theta d\phi dv \\ \notag
&= 3\sqrt{3} \pi M^2 v.
\end{flalign}

Therefore, we see that in classical general relativity, an old Schwarzschild black hole has a larger interior volume compared to a young one, even if they have the same mass. Even more surprisingly, the volume continues to grow even if the black hole is losing mass via Hawking evaporation \cite{1503.08245,1604.07222}. A remnant therefore has a vast interior volume, although this is not visible from the outside. Thus, even if we can create mini black holes with powerful high energy collision, these newly formed black holes are \emph{not} the same as a remnant. 

As argued in \cite{2407.09584,1710.00218}, black holes may harbor more states than those surface states that account for the Bekenstein-Hawking entropy, and as the black hole shrinks to Planckian size remnant, the information remains inside until finally (after a lifetime that scales as $M^4$) the remnant decays into a large number of low energy photons. The huge number is required if they were to carry all the information inside. In such a scenario, the usual Hawking radiation need not carry any information. In order to produce a remnant, one has to consider essentially the reverse process: focusing a large number of low energy photons. This explains why it is actually very difficult to form a remnant in high energy physics experiments, or even spontaneously from the vacuum. This is similar to the results in \cite{1108.0417}, in which it was argued that the most likely history of a fluctuation out of equilibrium is simply the CPT conjugate of the most likely way a system relaxes back to equilibrium. This means that though given enough time, systems near equilibrium can fluctuate into lower-entropy states, it is exceedingly unlikely to fluctuate out an egg; the most likely way to get an egg formed from fluctuation is for the shards of egg shells and runny yolk and egg white to gradually formed and re-assemble into an egg, i.e., the reverse process of breaking an egg. In general, the most likely way to get into a low entropy state is via a sequence of individually unlikely events. For our case, a black hole remnant decays into a large number of photons at the end of its long lifetime. By the second law of thermodynamics, the collection of photons must have a higher entropy state than the remnant itself. Thus, the most likely way for a remnant to be spontaneously created would be for a large number of low energy photons to somehow gather together by chance. This itself is already exceedingly unlikely to happen, so to spontaneously produce a remnant directly from fluctuation is even more unlikely.

In order for remnants to be viable, one has to also address the \emph{entropy problem}. As previously mentioned, the question was how does a ``small'' remnant possibly store enough entropy. We have already seen that the remnant is not small, quite to the contrary. However, the entropy problem also concerns what is the nature of the Bekenstein-Hawking entropy, namely whether it reflects all the degrees of freedom inside the black hole. If so, then it does not matter how large the interior is, it cannot contain that much information for some reason. This is the position of the ``central dogma'' -- that the number of states inside of black hole goes like $e^{A/4}$. The issue of the interpretation of the Bekenstein-Hawking entropy is of course an old problem (see, e.g., \cite{0501103}), but the view that the Bekenstein-Hawking entropy measures all the available states (the ``strong form'' interpretation \cite{0901.3156}) became a dominant view due partly to the success of holography (AdS/CFT correspondence). Still, it could be that the Bekenstein-Hawking entropy measures only the ``surface states'' (the ``weak form'' interpretation \cite{0901.3156}) that are relevant for some physical purposes, while the total number of states inside the black hole far exceeds $e^{A/4}$. The area is relevant for black hole thermodynamics, and by extension holography, but there can be other ``non-thermodynamic states'' inside.
Indeed, as mentioned in \cite{2407.09584} the entanglement (von Neumann) entropy is only guaranteed to be equal or smaller than the thermodynamic entropy for ergodic systems, but black holes are not ergodic. 

There are other subtleties that one should consider. As mentioned, the large volume inside a black hole is crucial for understanding a remnant. An intuitive picture can also be found in \cite{2407.09584}, namely that the mass of the black hole is essentially its energy. If the remnant is Planckian, it has very low energy, thus in order to account for huge entropy (large number of states), we need a lot of low energy, i.e., long wavelength particles, inside. This can only be possible if the interior volume is huge, which is indeed the case. Thus, treating remnants as point particles cannot possibly work from the onset, and it is that assumption that led to the idea of remnants being prematurely rejected. Another issue, as already pointed out in \cite{0901.3156}, is that the species problem rests on an effective field theory argument (which is required to estimate the production cross section), which in turn means that there is a cutoff scale $\lambda_c \gg L_p$ above which the remnant is treated as a point particle. However, due to the highly curved spacetime, it is nontrivial to define such an effective field theory. Indeed, due to gravitational redshift/blueshift, a longer wavelength far away from the black hole, is short close to it, and long again in the interior, and hence a fixed cutoff does not seem to make much sense. This is not to say that we cannot have an effective field theory on curved spacetime at all, but it would be too premature to rule out remnant based on an argument that is itself rather non-rigorous. 

Another subtlety concerns the notion of entropy itself being observer-dependent (which is not really a surprise because it is well-known that temperature is also observer-dependent -- the Hawking temperature being valid for asymptotic observers). In particular, consider a wave packet of fixed energy in flat spacetime. We can increase the entropy without adding more energy by introducing more particle species. Suppose there are some huge number $n$ of free scalar fields. Then the entropy grows like $\log n$. If we go to the Rindler wedge instead, then the von Neumann entropy of the mixed state, though increases as $\log n$, is bounded above by $E/T$ \cite{0310022} (note that this is essentially the Bekenstein bound, up to a constant prefactor). This at least suggests that we should be careful about \emph{whose} entropy are we talking about. Perhaps the interior of a remnant can have vastly more entropy than its surface bound, without being in tension with the fact that an asymptotic observer only sees the surface entropy. In addition, even the interpretation of the Bekenstein bound requires some care; in \cite{2309.07436} Hayden and Wang argued that unlike classical bits and qubits, zero-bits\footnote{A zero-bit is a special case of $\alpha$-bits, introduced in \cite{1706.09434}. They correspond to the ability to perform quantum error correction on arbitrary bounded dimension subspaces, with the real number $0 \leqslant \alpha \leqslant 1$ characterizing the size of the subspace. The standard quantum error correction corresponds to $\alpha=1$, while $\alpha=0$ is related to quantum identification, a form
of equality testing for quantum states. In the context of black holes, if the central dogma holds, the zero-bits are revealed after the Page time \cite{1807.06041}. The information scrambling of Hayden-Preskill protocol \cite{0708.4025} also corresponds to the $\alpha=0$ case \cite{1807.06041}.} and their associated information processing capability are generally not constrained by the Bekenstein bound. In other words, we must understand the various subtleties of entropy before we can definitely rule out remnants.

\section{Discussion: Prospects for Black Hole Remnants}

As we have seen, there are various ways one could obtain black hole (or white hole) remnants. It should be emphasized that the GUP approach, though relatively simple, still requires more careful investigations \cite{2005.12075,2303.10719,2305.16193}. After all, there are many versions of GUP in the literature, which should we trust? Even for a fixed GUP, it is not clear what should the approach be for black hole thermodynamics. For example, should we modify only the Hawking temperature and the entropy, or also modify the metric tensor? If the latter, how do we obtain the metric? No unique, well accepted, answers are available to these questions in the literature \cite{2303.10719}, and as such the conclusions obtained about remnants are also subject to a certain degree of skepticism. The possibility that black holes can quantum tunnel into a white hole remnant, obtained mostly from loop quantum gravity approach, seems more promising.

In principle remnants can still ``resolve'' the information paradox, but one has to be careful of some subtle aspects. For one, whether there is a remnant or not is quite independent from whether Hawking radiation encodes information. If information is still carried out via Hawking emission, then the firewall paradox remains because Page time sets in much earlier than the remnant stage \cite{1412.8366} if the remnant is Planckian. However, the requirement that information be carried out by Hawking radiation rests upon the assumption of the central dogma, which as we have discussed, need not hold. Therefore, Hawking radiation can be information free like initially thought, with the information somehow contained within the large interior of the remnant. If the remnant slowly emits diffuse particles and completely evaporates in a far longer time scale (at least $M^4$), then eventually all the information is released back to the exterior universe and unitarity is preserved. The large interior allows these particles to be soft particles that carry little energy, which also explains why remnants are not copiously produced in everyday high energy experiments -- any mini-black hole we can produce in future laboratory is a completely different object from a remnant. Much of the discussions here are not new, unfortunately the current state remains more or less the same, with many colleagues still unaware of the counter-arguments against the species and entropy problems. 

The interior volume of black holes did however, gained some attention in the high energy community due to it possibly being a holographic dual to complexity (the so-called ``Complexity-Volume conjecture'' or just CV-conjecture \cite{1402.5674,1403.5695,1406.2678,1411.0690,1509.06614,1911.12561,2312.05731}), yet how this might relate to a remnant is still unclear. In \cite{1507.02287}, Susskind proposed the idea that black hole complexity cannot grow without bound. Instead, after some time that scales as $\exp{(S)}$, where $S$ is the Bekenstein-Hawking entropy\footnote{This time scale, exponential in the entropy, is also the time scale that is required for one to decode the information contained in Hawking radiation \cite{1301.4504}, assuming the latter contains information.} , the quantum state of the black hole now also contains a significant white hole component. This is the ``gray hole'' phase\footnote{A different kind of ``gray hole'' is considered in \cite{2853916}.}, during which the complexity fluctuates around a constant value. After a time of $t\sim \exp[{\exp{(S)}}]$ the ``gray hole'' becomes a white hole, during which the complexity decreases. Note that this is very similar to the black-to-white hole transition we have discussed, but the time scale involved is extremely different. The possibility that even astrophysical black holes are ``gray'', i.e., in the quantum superposition of black and white hole states has also been discussed in \cite{2304.10692}, in which the authors also pointed out it is possible to arrange the ``black-to-white transition'' to have zero action -- so that it will not be subject to destructive interference in the Feynman path integral.
 
We should also not rule out the possibility that information escapes into ``another universe'', interpreted broadly. For example, many of the regular black hole solutions have de-Sitter core or other geometries (for example, an interior which asymptotes to $dS_2 \times S^2$ in the future is obtained in \cite{2005.13260}), which essentially is like another universe whose spacetime dimension becomes effectively 2-dimensional (the size of the $S^2$ is fixed). Even under the usual Hawking evaporation without remnant, we already mentioned that the interior volume continues to grow. Therefore, if the black hole completely evaporates away, the interior spacetime must pinch off its parent (our universe) and subsequently becomes a new baby universe, taking away all the information with it (assuming that Hawking radiation is information free) \cite{9405007v2}. The idea that black holes lead to a new universe and thus avoiding the information loss problem is of course, not new either; see, e.g., \cite{0103019}. The pinching-off of a baby universe is sometimes referred to as the black hole ``emitting'' a baby universe. In \cite{2208.01625}, based on a calculation carried out in Jackiw-Teitelboim (JT) gravity, it was argued that a very old black hole can tunnel to a white hole or even a firewall by emitting a large baby universe. However, the probability for tunneling to a white hole is proportional to $t^2\exp{(-2S)}$, where t is the age of the black hole, which again differs from the loop quantum gravity tunneling probability of $\exp{(-S)}$ \cite{2407.09584} (granted that the gravity theories are different). In addition, the loop quantum gravity tunneling does not ``emit'' away the large interior; the interior just slowly shrinks throughout the white hole phase until the remnant completely disappears via diffuse emission.

It is also interesting to compare the remnant picture with that of the usual remnant-free end stage of Hawking evaporation. Note that as a Schwarzschild black hole evaporates, it gets smaller, which in turn means that the curvature (say, the Kretschmann scalar) at and near its horizon is becoming larger. In a truly quantum picture $M$ cannot be continuously changed (although semi-classically we model the evaporation with a differential equation of $dM/dt$), so at the final stage the Planckian black hole is expected to dissolve away into a collection of particles and we are left with a flat spacetime. Still, this is only an expectation. How do we know that the large spacetime curvature does not ``tear'' spacetime enough to lead to a naked singularity? Certainly such a possibility has been hinted at in \cite{russo, hiscock}.
Including higher curvature terms (quadratic gravity) also suggests such a possibility \cite{2409.16690}, though it is hard to say what the full quantum gravity result would be. 
We normally expect that the end state is flat spacetime because it is the smooth limit of $M \to 0$, but it is \emph{not} a smooth limit for the curvature. Of course, naked singularity as the end state will be bad, and by the (weak) cosmic censorship conjecture it is natural to think that this does not happen. Perhaps remnants also play the role of a cosmic censor. Still, this does not completely avoid the problem, as we will discuss below.

In the remnant picture, if the lifetime is finite, the same thing happens towards the end. We do not know for sure whether quantum gravity can indeed ``resolve'' singularity completely\footnote{Note that GUP may or may not remove the classical singularity. The remnant obtained in \cite{0106080} still considers the same classical form of Schwarzschild geometry, so the singularity is untouched (here the ``minimal length'' is reflected in the size of the remnant, not in ``smearing out'' the singularity. For a review about minimal length, see \cite{1203.6191}.). Though in other models, one could also employ GUP to obtain regular black holes \cite{1305.3851,2109.05974}.}, or replace it with a milder, quantum version (see \cite{2005.07032,2210.16856} for my related reviews). In fact, if a remnant has more entropy than the Bekenstein-Hawking entropy, then a version of singularity theorem can be proved \cite{2201.11132}. 
This means that it is still possible that the end state is either a singularity or something else that can only be described by quantum gravity. As Penrose wrote \cite{RP}: 
\begin{quote}``It
is hard to avoid the conclusion that the endpoint of the Hawking
evaporation of a black hole would be a naked singularity --- or at
least something that one [sic] a classical scale would closely resemble
a naked singularity.'' \end{quote}
So both the standard picture and the remnant picture may encounter singularity at the end if they are not sufficiently resolved in quantum gravity. 

Even if the lifetime of the black hole is infinite, one cannot avoid this problem as long as the mass becomes Planckian. This is because naked timelike or null singularities are bad in classical general relativity not because of the singularities themselves, but because their presence renders the theory non-predictive. Without knowing the properties of the singularities, one cannot know what influences it can have on spacetime region even far away from them. Note that this problem does not only arise when the curvature is mathematically divergent, it already arises for large enough curvatures. Usually we expect general relativity to be a good theory when the curvature scale is above the Planck scale (for the subtleties, see \cite{2303.10719}). Thus, if black holes become Planckian in size (whether there is a Planck size remnant or not), cosmic censorship will be violated in spirit. Again, it is worth emphasizing that stable remnants (those with infinite lifetime) cannot serve as a cosmic censor, it effectively becomes a ``Planckian curvature source'', as bad as a naked singularity. This is not necessarily a disaster, it just means that the end state of Hawking evaporation cannot be understood semi-classically. Nevertheless, we should keep in mind that one of the criticisms against firewall is that it is essentially a naked singularity \cite{1511.05695}. Should we prefer remnants over firewalls if they have the same problems? Perhaps the former is more palatable because it only arises for Planckian black holes, whereas the latter can occur even for large black holes. 

Remnants that stop evaporating after either the Page time or scrambling time is effectively classical and this issue never arises. One might think that this problem also does not arise for extremal black hole remnants if they asymptote to extremal state when still being macroscopic in size. However, this is not the case, as various studies have suggested that near-extremal black holes are highly quantum object and thus the result cannot be trusted \cite{0209039,1005.2999,2210.02473,2303.07358,2307.10423,2309.04110,2409.08236}. In fact, an extremal black hole can be effectively a singular object \cite{0209039,2210.02473,2303.07358}.

To conclude, no argument against remnants holds up to close scrutiny, and thus the option that remnants can avoid information paradox remains viable. Given that Planck size remnants may also serve as a dark matter candidate (see, e.g., \cite{MacGibbon,0205106,0406514,1805.03872}) --- a possibility already noted by  Aharonov,
Casher, and Nussinov \cite{ACN} --- it would be interesting to further investigate their properties under different models. For example, recently there was a concern that some asymptotically extremal black hole remnants may be too fast (due to the recoil of the emitted radiation) \cite{2102.06517} and thus too warm to be dark matter, but subsequent discussions had reassured that it is still viable \cite{2104.08919,2105.01627}. 

In the complete evaporation picture, the lower bound of the primordial black hole mass is set by the Hawking evaporation rate, namely that black holes cannot be too small or otherwise they would have completely evaporated by now. This yields a lower bound of about $10^{17}$ g. With remnants, this bound has to be modified, depending on the model. For example in \cite{2407.09584}, the white hole remnant is estimated to be in the mass range of $10^{10}$ g and $10^{15}$ g, and Schwarzschild radius in the range $10^{-18}$ cm and $10^{-13}$ cm. These bounds are obtained from the requirement that the white hole remnant lifetime is longer than the Hubble time, and its progenitor black hole's lifetime is shorter than the Hubble time, namely $M^3 < t_\text{Hubble} < M^4$. If, however, black holes already stop evaporating after the Page time or even the scrambling time, then again we must re-check the bounds. In \cite{2402.17823}, it is shown that if the remnant forms after about half of its mass is evaporated (the ``memory burden model''), then we could have light primordial black holes below $10^{10}$ g serving as a dark matter candidate. It has recently been pointed out that the evaporation rate for cosmological black holes cannot be treated the same way as isolated black holes in asymptotically flat spacetimes. The consequence is that the decay of primordial black holes occurs faster for larger masses but the decay rate reduces for lower mass \cite{2110.14379}.  
Some mass ranges of remnants can hopefully be tested in the future via gravitational wave signature by, e.g., the Einstein Telescope. One such scenario was studied in \cite{2303.07661}, in which it was pointed out that if the remnant mass is around $5 \times 10^5$ g, then there will be a cosmological gravitational wave signal at frequencies around 100 Hz. The aforementioned memory-burden proposal can potentially also be tested via gravitational waves \cite{2409.06365}.  What is the actual mass scale for a black hole transitioning into a remnant is therefore an important question that requires more studies.

%%%%%%%%%%%%%%%%%%%%%%%%%%%%%%%%%%%%%%%%%%%%%%%%%%%%%%%%%%%%%%%%%%%%%%%%%%%%%%%%%%%%%%%%%%%%%%%%%%%%%%%%%%%%%%%%%%%%%%%%%%%%%

\end{document}